\documentclass[11pt,twoside]{article}

\usepackage{asp2006}
\usepackage{epsf}
\usepackage{lscape}

\usepackage{graphicx}

\begin{document}
\title{Invisible Plasma Content in Blazars? 
The Case of Markarian 421}   
\author{M. Kino}   
\affil{Institute of Space Astronautical Sicnce, JAXA, 3-1-1 Yoshinodai, 
Sagamihara, Kanagawa 229-8510, Japan}    
\author{F. Takahara}   
\affil{Department of Earth and Space Science, Osaka University, 
560-0043 Toyonaka, Japan\\}    

\begin{abstract} 

Invisible plasma content in blazar jets such as protons and/or thermal 
electron-positron ($e^{\pm}$) pairs is explored through combined 
arguments of dynamical and radiative processes. By comparing physical 
quantities required by the internal shock model with those obtained 
through the observed radio-to-gamma-ray spectra for Mrk 421, we find  
the existence of a copious amount of invisible plasma in the jet. 
We speculate that the blazar sequence could arise from 
variations of total amount and/or blending ratio 
of $e^{\pm}$ pair and electron-proton plasma. 

\end{abstract}


\section{Introduction}

The plasma content of relativistic jets is not 
easily constrained by observations since the emission is 
dominated by that from non-thermal electrons and
it is difficult to directly constrain thermal plasma. 
Hence, the plasma composition in AGN jets, whether normal 
proton-electron ($e/p$) plasma 
or electron-positron pairs ($e^{\pm}$) is 
a dominant composition, is 
still a matter of debate
(e.g., Reynolds et al. 1996; 
Wardle et al. 1998; Hirotani et al. 1999;
Sikora \& Madejski 2000). 
This problem prevents us from estimating
the total mass and energy flux ejected from a
central engine.
In order to constrain the invisible matter in the jets, 
dynamical considerations are indispensable.
In Kino \& Takahara (2004), we proposed a new procedure  
to constrain the invisible plasma component
in classical FR II radio sources.
We used the fact that the mass and energy densities of 
the sum of thermal and non-thermal (NT hereafter) particles 
are larger than those of non-thermal electrons which 
are determined by observations.
We apply the same technique to the inner core jets of 
AGNs (i.e., blazars) based on the standard internal shock model
(Kino \& Takahara 2008).

\section{Mass density of invisible plasma}

The methodology of constraining 
the invisible plasma content in the emission region is as follows.
A lower limit to the total mass density (sum 
of non-thermal electrons and invisible plasma)  
is restricted by the definition that the mass density
of total plasma ($\rho$) should be smaller than that of 
the non-thermal electrons ($\rho^{\rm NT}_{e}$). 
Similarly, 
a lower limit to the total internal energy density ($e$)  
is limited by the definition that $e$ is smaller than 
that of the NT electrons ($e^{\rm NT}_{e}$). 
Using a relativistic shock jump conditions,
we can estimate the values of $\rho$ and $e$ 
(see Kino \& Takahara 2008).
The mass density of non-thermal electrons of TeV blazars
can be well estimated by the fitting of a
multi-frequency spectrum with synchrotron self-Compton (SSC) model 
(Kino et al. 2002, hereafter KTK).
The number density of non-thermal electrons $n_{e}^{\rm NT}$  
has been estimated as
$n_{e}^{\rm NT}\equiv
\int^{\infty}_{\gamma_{e,\rm min}}
n_{e}(\gamma_{e})d\gamma_{e}$,
while $e_{e}^{\rm NT}$ is given by
$e_{e}^{\rm NT}=
\langle\gamma_{e}\rangle
n_{e}^{\rm NT}m_{e}c^{2}$, 
where 
$n_{e}(\gamma_{e})$ and 
$\langle\gamma_{e}\rangle$ are the energy spectrum and  
the average Lorentz factor of NT electrons, respectively.
By a detailed comparison of the SSC model with the
observed broadband spectrum of Mrk 421,
we obtained $n_{e}^{\rm NT}$ as
$n_{e}^{\rm NT}\simeq
11\times (\gamma_{e,\rm min}/10)^{-0.6} \ \rm cm^{-3}$.
Here we adopt the index of injected electrons for Mrk 421 
as $s=1.6$ (e.g., Kirk \& Duffy 1999).
As for the average energy of NT electrons, we obtained  
\begin{eqnarray} \label{eq:uacc}
e_{e}^{\rm NT}/n_{e}^{\rm NT}
=\langle\gamma_{e}\rangle m_{e}c^{2}
\simeq
3.1\times 10^{2} m_{e}c^{2} ,
\end{eqnarray}
where $\langle\gamma_{e}\rangle$ is the average
Lorentz factor of 
NT electrons.
Since for $s=1.6$,  
electrons near the cooling break energy $\gamma_{e,\rm br}\approx10^{4}$
carry most part of the kinetic energy
and $e_{e}^{\rm NT}$ has a weak dependence on 
the minimum Lorentz factor $\gamma_{e,\rm min}$.
In Fig. 1, we show that the case of 
$\gamma_{e,\rm min}\approx 10^{4}$ is ruled out for Mrk 421
since the case does not fit the EGRET data (KTK).
Therefore, Eq. (\ref{eq:uacc}) is justified in any case 
for Mrk 421.

The condition of 
$e_{e}^{\rm NT}/e
=\langle\gamma_{e}\rangle\rho_{e}^{\rm NT}
/\Gamma_{\rm rel}\rho<1$ can be rewritten as 
\begin{eqnarray}\label{eq:lower}
\frac
{\rho} 
{\rho_{e}^{\rm NT}}
&>& \frac
{\langle\gamma_{e}\rangle}
{\Gamma_{\rm rel}}
\simeq \frac{3.1\times 10^{2}}{\Gamma_{\rm rel}} ,
\end{eqnarray}
where $\Gamma_{\rm rel}$ is the relative bulk Lorentz factor
between an upstream and a downstream of the shock.
From this, we directly see that
the invisible mass density at least about 100 times the rest mass density 
of NT electrons is definitely required
in the framework of internal shock model in which 
$\Gamma_{\rm rel}\le a  few$ takes place. 
In other words,
a loading of baryons and/or a thermal pair plasma is expected. 
As for the upper limit,
we use the constraint that bremsstrahlung emission from 
thermal electron (and positron) component should 
not exceed the observed $\gamma$-ray emission. 
We can thus bracket the amount of total mass density
in the emission region from below and above. 
In this way we apply this method to
the TeV blazar Mrk 421.

In Fig. 2 we show the resultant $\rho/\rho_e^{\rm NT}$
plotted versus $\Gamma_{\rm r}/\Gamma_{\rm s}$
in the cases of ``equal $\rho$'' and 
``equal mass''  ($m=\rho \Gamma\Delta$)
where 
$\Gamma_{\rm r}$,
$\Gamma_{\rm s}$, and
$\Delta$ denote
the Lorentz factors of rapid and slow shell, and
the thickness of the shell 
measured in ISM frame, respectively. 
We impose the condition $\Delta_{\rm r}/\Delta_{\rm s}=1$ 
(e.g., Spada et al. 2001).
As for the plasma content, we examined the two simple cases.
One is the case of the jet 
with pure $e^{\pm}$ pair plasma content,
whilst the other is the jet made of pure $e/p$ plasma. 
The qualitative features are the same for these two cases, 
although different in quantitative detail.
Central to the results is that a
large amount of mass density of invisible plasma 
is clearly required in the emission region.
As the value of $\Gamma_{r}/\Gamma_{s}$ increases, 
the value of $\Gamma_{\rm rel}$ becomes larger
and the lower limit on the invisible mass density 
($\rho/\rho^{\rm NT}_{e}$) 
reduces. 


\section{Discussion}

Let us  discuss 
the origin of blazar sequence (Fossati et al. 1998)
which is tightly connected to the nature of the central engine.
In Fossati et al. (1998)
they computed average spectral energy distributions from radio 
to gamma-rays for complete sample of blazars.
They claimed that the continuous sequence of properties  
may be controlled by a single parameter, related to 
the bolometric luminosity.
Here we enlighten another new ingredient for the origin of sequence.
Flat spectrum radio quasars
(FSRQs) have the order of magnitude larger kinetic power and 
larger energy densitity of the surrounding external radiation field 
than BL Lacs (e.g., Sikora et al. 1997). 
Hence the leptonic components in FSRQs ejecta 
undergo severe radiation drag
in dense external radiation fields 
(Sikora \& Wilson 1981; Phinney 1982).
However, the bulk Lorentz factors in FSRQs are
slightly larger than the ones in TeV blazars 
(Jorstad et al. 2001; Kellermann et al. 2004)
in spite of being subject to much stronger radiation drag
(e.g., Kubo et al. 1998; Spada et al. 2001).
In order to realize the larger kinetic powers 
and larger bulk Lorentz factors against the strong radiation drag, 
we may take a new conjecture that 
a larger baryon loading may occur for FSRQs.
Summing up,  not only the strength of the external radiation field
but also the total amount and/or blending ratio 
of $e^{\pm}$ pair and $e/p$ could be new key 
quantities to explore the origin of the continuous blazar sequence.

\begin{figure} 
\begin{center}
\includegraphics[width=6.2cm]{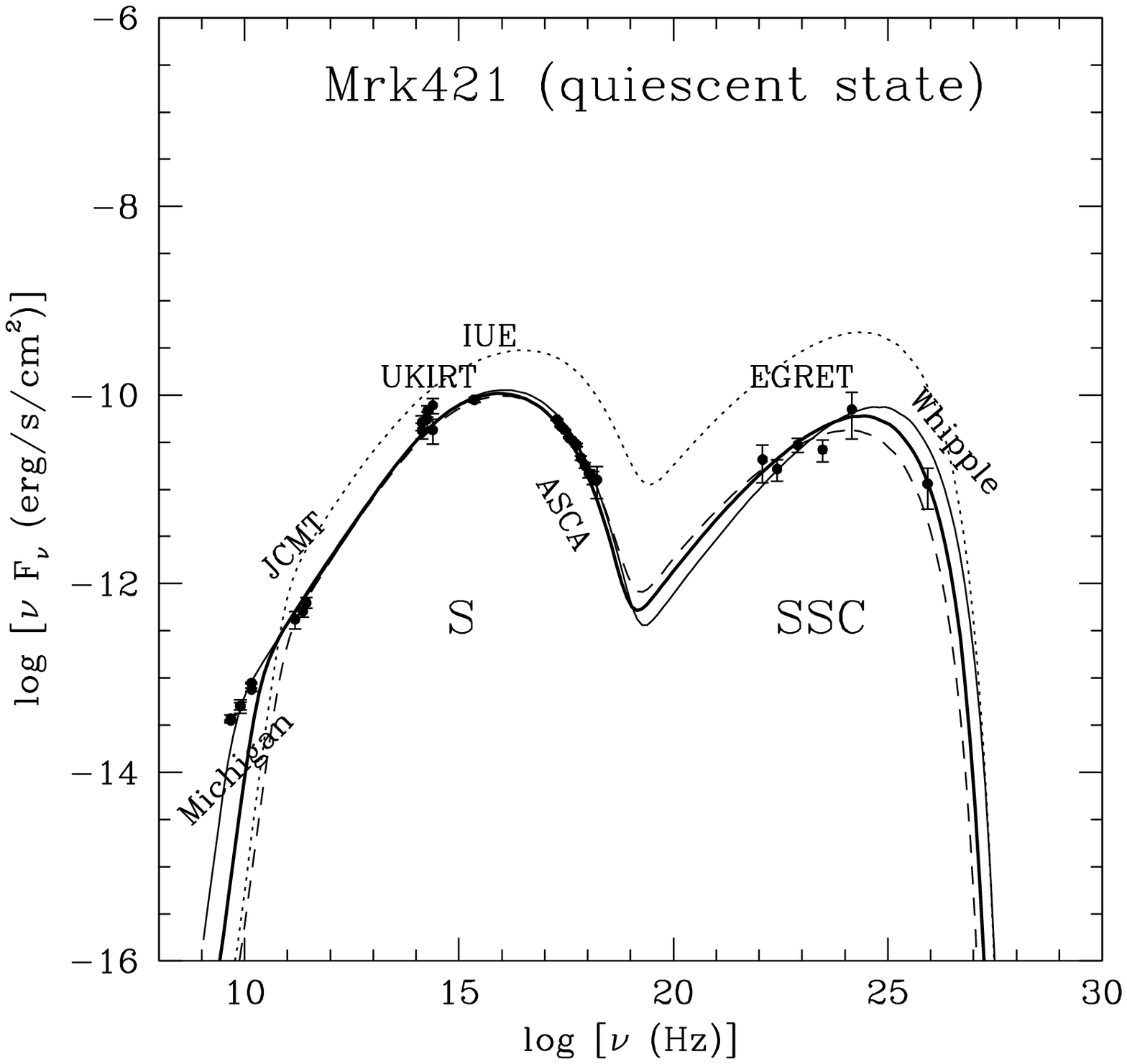}
\includegraphics[width=6.2cm]{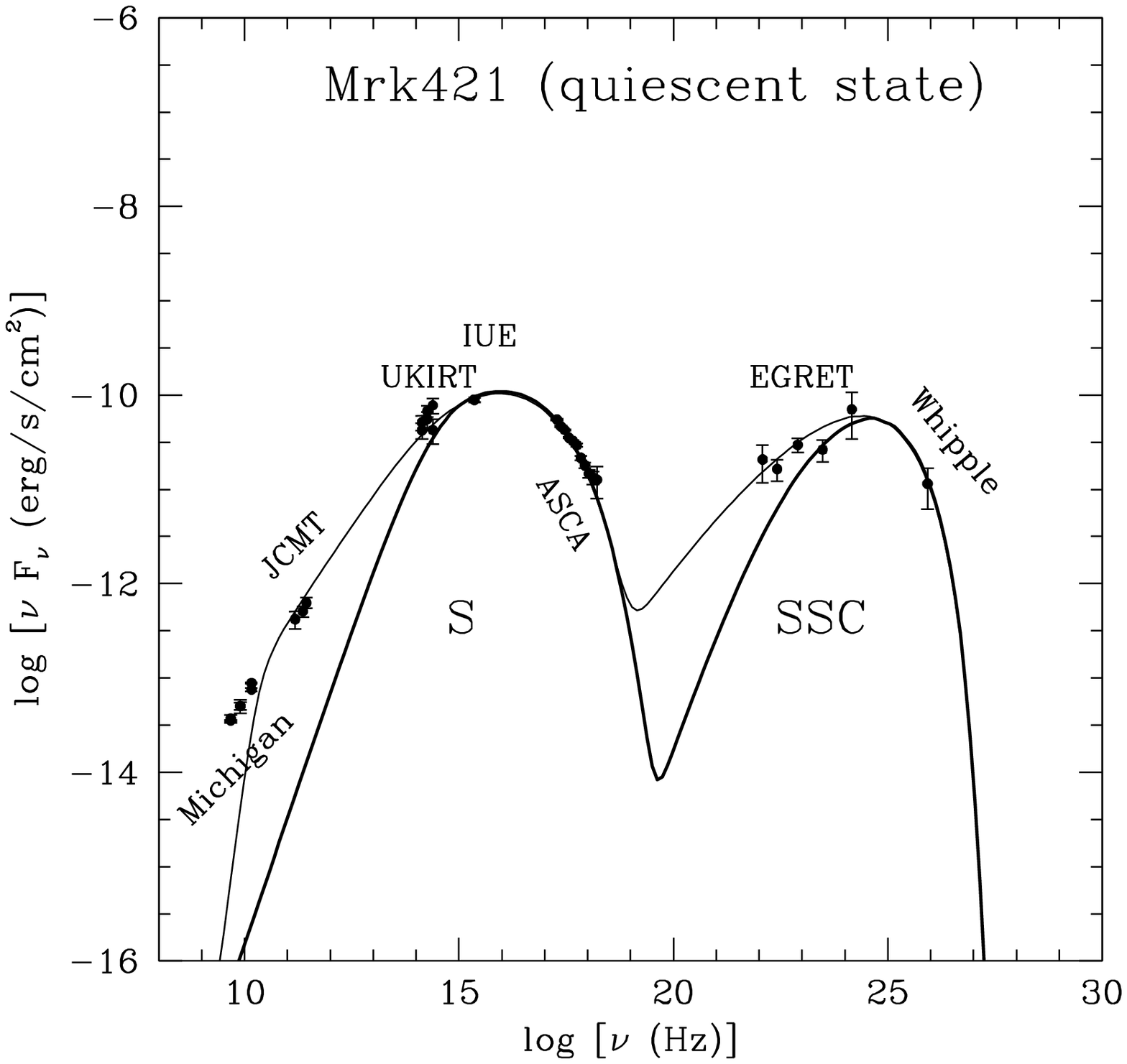}
\end{center}
\caption
{One-zone SSC model spectra for the steady state emission of Mrk 421
(KTK).
The thick solid line in the left one shows the best fit spectrum.
In the right one, we examine the case of 
$\gamma_{e,\rm min}=10^{4}$ (the thick solid line). 
The case of $\gamma_{e,\rm min}=10^{4}$ is ruled out because 
it cannot fit the spectrum well.}
\label{fig:eqmass}
\end{figure}
\begin{figure} 
\begin{center}
\includegraphics[width=6.5cm]{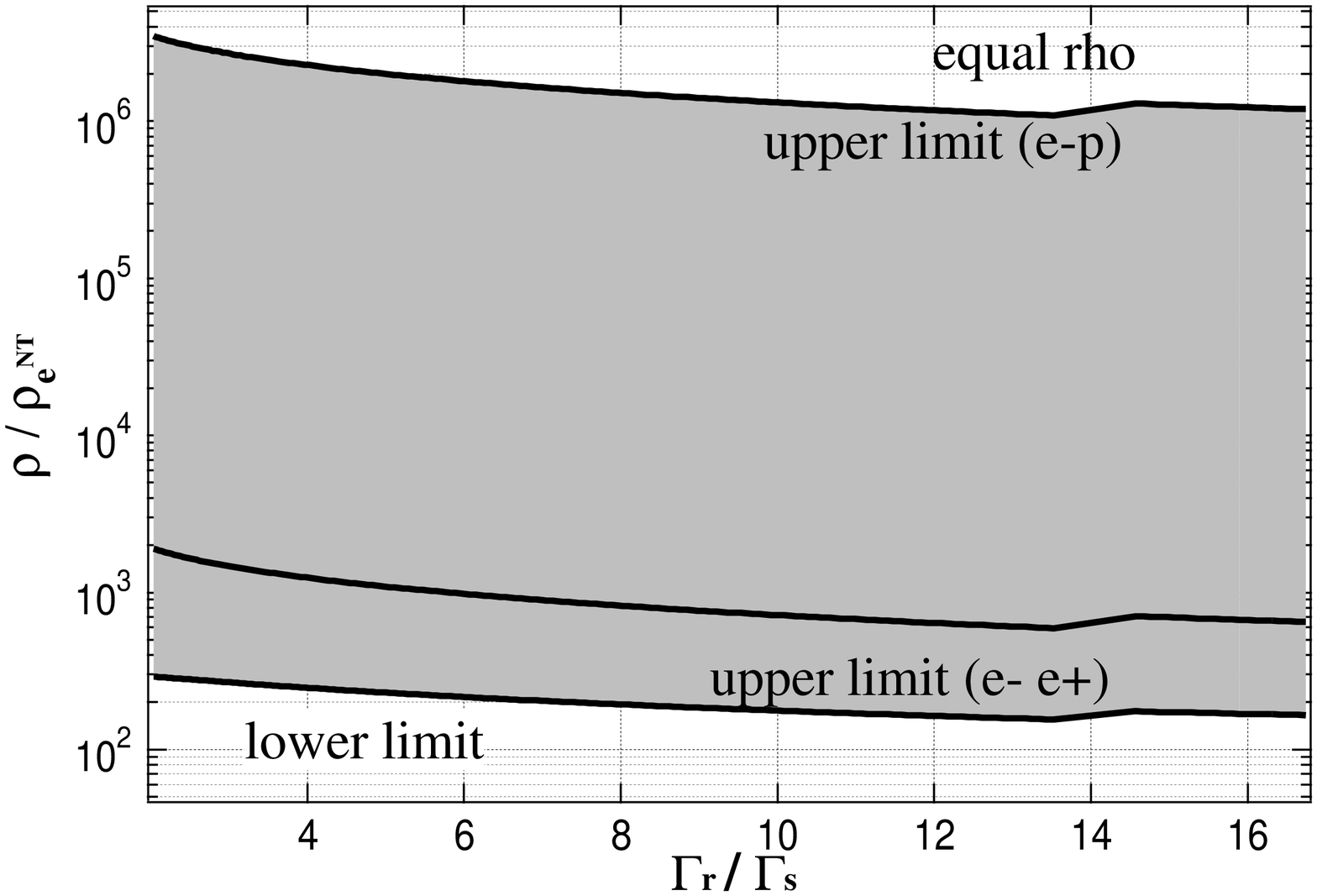}
\includegraphics[width=6.5cm]{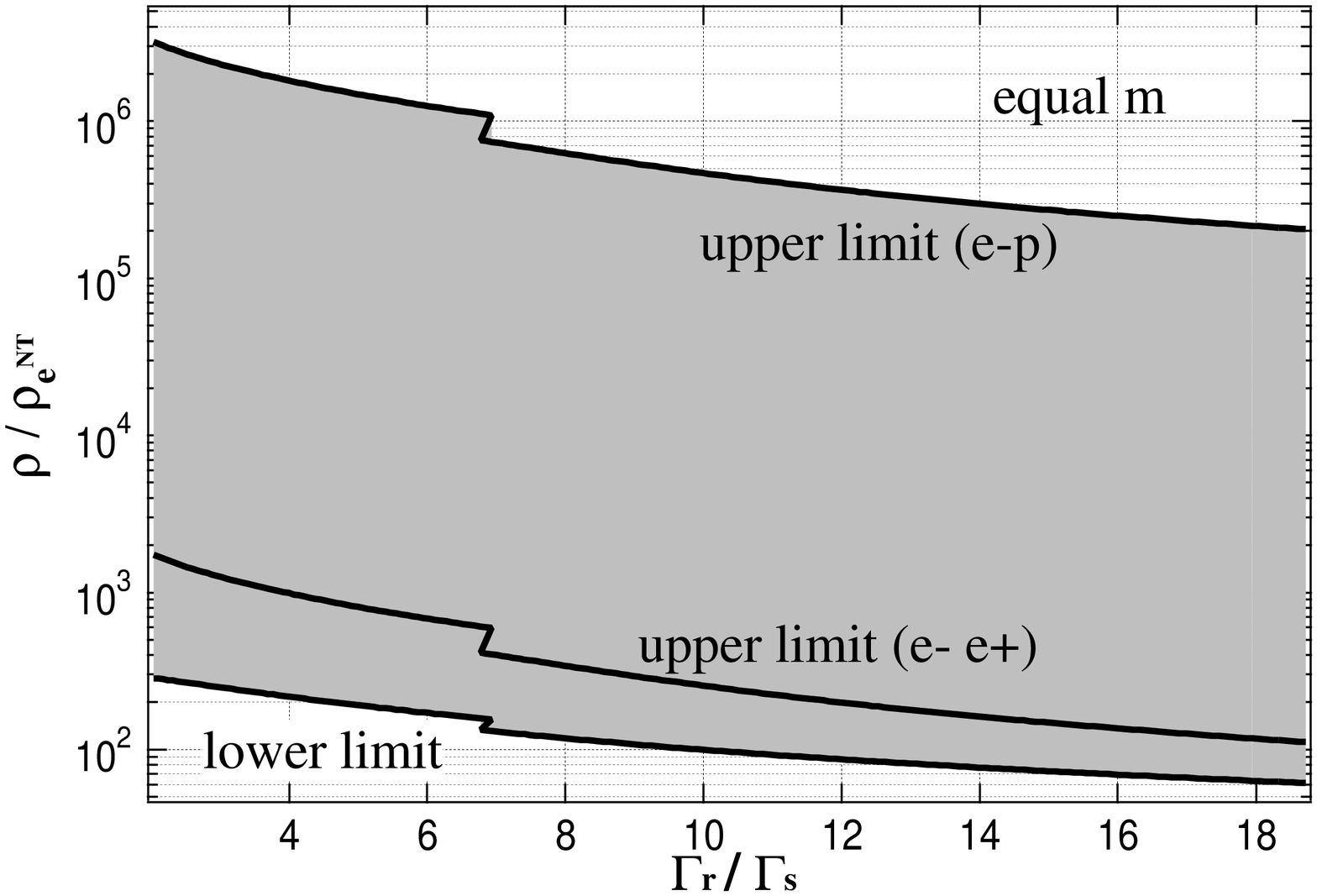}

\end{center}
\caption
{The allowed regions (the gray shaded regions) 
of $\rho/\rho_{e}^{\rm NT}$;
the amount of mass density of total plasma
normalized by that of non-thermal electrons
for ``equal $\rho$'' (left) and  ``equal $m$'' (right). 
Horizontal axis shows the ratio of the Lorentz factor
of a rapid shell to a slow one.  }
\end{figure}

\acknowledgements 

We acknowledge the
Grant-in-Aid for Scientific Research 
of the Japanese Ministry of Education, Culture, Sports, Science
and Technology, No. 14079025, 14340066, and 16540215.


\end{document}